\documentclass[twocolumn]{aastex61}
\received{... ..., 2017}
\revised{... ..., 2017}
\accepted{... ..., 2017}
\submitjournal{ApJ}

\shorttitle{}
\shortauthors{Shen et al.}
\newcommand{\speed}[1]{#1 km~s${}^{-1}$}
\newcommand{\accel}[1]{#1 m~s${}^{-2}$}
\newcommand{\nfig}[1]{Figure~\ref{#1}}
\newcommand{\rsun}[1]{${#1}\,R_\odot$}

\begin{document}

\title{On a small-scale EUV wave: the driving mechanism and the associated oscillating filament}
\correspondingauthor{Yuandeng Shen}
\email{ydshen@ynao.ac.cn}
\author{Yuandeng Shen}
\affiliation{Yunnan Observatories, Chinese Academy of Sciences,  Kunming, 650216, China}
\affiliation{State Key Laboratory of Space Weather, Chinese Academy of Sciences, Beijing 100190, China}
\affiliation{Center for Astronomical Mega-Science, Chinese Academy of Sciences, Beijing, 100012, China}
\author{Yu Liu}
\affiliation{Yunnan Observatories, Chinese Academy of Sciences,  Kunming, 650216, China}
\affiliation{Center for Astronomical Mega-Science, Chinese Academy of Sciences, Beijing, 100012, China}
\author{Zhanjun Tian}
\affiliation{Yunnan Observatories, Chinese Academy of Sciences,  Kunming, 650216, China}
\author{Zhining Qu}
\affiliation{School of Physics and Electronic Engineering, Sichuan University of Science \& Engineering, Zigong 643000, China}

\begin{abstract}
We present observations of a small-scale Extreme-ultraviolet (EUV) wave that was associated with a mini-filament eruption and a {\em GOES} B1.9 micro-flare in the quiet Sun region. The initiation of the event was due to the photospheric magnetic emergence and cancellation in the eruption source region, which first caused the ejection of a small plasma ejecta, then the ejecta impacted on a nearby mini-filament and thereby led to the filament's eruption and the associated flare. During the filament eruption, an EUV wave at a speed of \speed{182 -- 317} was formed ahead of an expanding coronal loop, which propagated faster than the expanding loop and showed obvious deceleration and reflection during the propagation. In addition, the EUV wave further resulted in the transverse oscillation of a remote filament whose period and damping time are 15 and 60 minutes, respectively. Based on the observational results, we propose that the small-scale EUV wave should be a fast-mode magnetosonic wave that was driven by the the expanding coronal loop. Moreover, with the application of filament seismology, it is estimated that the radial magnetic field strength is about 7 Gauss. The observations also suggest that small-scale EUV waves associated with miniature solar eruptions share similar driving mechanism and observational characteristics with their large-scale counterparts.

\end{abstract}

\section{Introduction}
Global propagating magnetohydrodynamics (MHD) waves in the solar atmosphere were firstly observed in the dense solar chromosphere. They are historically called Moreton waves\citep[e.g.,][]{moreton60,moreton64,warmuth04a}, and are observed as arc-shaped bright propagating fronts at speeds of about \speed{1000} \citep[e.g.,][]{shen12a,shen14a,liu14,warmuth15}. \cite{uchida68} first proposed that   Moreton waves are the footprints of coronal shock waves in the dense chromosphere (see also \cite{vrsnak16}). \cite{moses97} and \cite{thompson98} first reported the similar wave-like phenomena (namely ``Extreme-ultraviolet (EUV)'' wave) in the solar corona. However, it is hard to consider EUV waves as the counterparts of Moreton waves due to their low speed (of \speed{200 -- 400} \cite{thompson09}), although the counterparts of Moreton waves have been detected at other heights of the solar atmosphere \citep[][]{khan02,vrsnak02,vrsnak05,kwon13}. Using high temporal and high spatial resolution observations taken by the Atmospheric Imaging Assembly \citep[AIA;][]{leme12}, \cite{nitta13} showed that the average speed of EUV waves is \speed{644}. However, this speed is still too much lower than that of Moreton waves. Recent observations indicate that the speed discrepancy between Moreton and EUV waves can be reconciled by considering the fast deceleration of EUV waves during the beginning stage \citep{warmuth04a,warmuth04b,shen12a,shen12b}. For example, \cite{shen12a} showed that a coronal wave and its corresponding chromospheric Moreton wave have a similar speed of \speed{1000} at the beginning stage, but the average speed of the coronal wave during the whole lifetime is only \speed{605}. So far, many scholar have accepted that EUV waves are the counterparts of Moreton waves in chromosphere.

During the past two decades, two main debating problems about EUV waves are about the driving source and the true physical property. Due to the lack of the knowledge of CMEs before 1970s, the driving source of Moreton waves was naturally considered to be the pressure enhancement of the flares \citep[e.g.,][]{moreton60,moreton64,uchida68}. Therefore, many earlier studies also believe that the driver of EUV waves are the associated flares \citep[e.g.,][]{warmuth04a,warmuth04b,cliver05}. However, along with the development of advanced solar telescopes, more and more studies based on high resolution observations  indicate that global EUV waves are in fact excited by the associated CMEs \citep[e.g.,][]{chen06,chen11,ma09,ma10,cheng12,shen12a,shen12b,shen13a,shen14a,muhr14,zhou17}. For the physical nature of EUV waves, many authors interpreted them as coronal compressive fast-mode magnetosonic waves \citep[e.g.,][]{thompson98,pats09a,muhr10,vero10,long11a,long11b,xue13,vanninathan15,yang13,long17}, while some peoples believe that EUV waves are not at all waves in physics \citep[e.g.,][]{dela99,fole03,harr03,attr09,attr10,zhuk09,dai10}. In most cases, only one bright wavefront can be detected. However, sometimes there are two simultaneous bright wavefront can be observed in a single eruption, in which one is fast and it can be detected in a large distance range, while the other is slow and its speed is about one third of the preceding fast one \citep[][]{dela99,dela00,chen11,shen12a,shen14a,kumar13,guo15}. \cite{chen16} recently found that part of a fast propagating wavefront can convert to a slow-mode wave, which is trapped inside magnetic loops and manifested as a stationary wave front. This phenomenon has benn confirmed by very recently  observations \citep{chandra16,yuan16,zong17}. In addition, recent high-resolution observations revealed that many large-scale EUV waves are tightly associated with the quasi-periodic fast-propagating magnotosonic waves that share common periods with the associated flares \citep[e.g.,][]{liu10,liu11,liu12,ofman11,shen12c,shen13b,yuan13,pascoe13,nistico14,kumar15,kumar17,goddard16}. On the other hand, solar physicists have made many theoretical efforts to understand the physical nature of EUV waves. For example, \cite{chen02} proposed a numerical model to interpret the co-existing fast and slow wavefronts in a single eruption, in which the authors explained that the fast wave is the corresponding coronal counterpart of the chromosphere Moreton wave, while the slow wavefront behind the fast one is formed by the successive opening of magnetic field lines. Therefore, the fast wavefront is a fast-mode magnetosonic wave in nature, but the slow one is not \citep{chen05}. The co-existing of both wave and non-wave explanation is supported by many observations \citep[e.g.,][]{zhuk04,cohe09,liu10,down11,li12b,li12c,shen12a,shen12b,shen13a,kumar13,zong15}. Other interpretations of EUV waves still include slow-mode wave \citep{will06,wang09,wang15}, successive reconnection model \citep{attr07,van08}, Joule heating mechanism \citep{dela07,dela08}, and so on. All these theoretical models can explain a part of characteristics of EUV waves. More information on the driving mechanism and physical property of EUV wave can be found in recent reviews \citep{gallagher11,patsourakos12,liu14,warmuth15}.

Previous studies mainly focused on large-scale EUV waves association with energetic solar eruptions. However, less energetic small-scale EUV waves associated with miniature eruptions should be also important in the full spectrum of EUV waves of hierarchic sizes. Although small-scale EUV waves are less energetic than their large-scale counterparts, they are more frequently and their total energy budget could be significant for the quiet-Sun \citep{liu14}. \cite{innes09} estimated that there are about 1400 miniature eruptions per day over the whole Sun, and about one third of them are associated with small-scale EUV waves that can propagate a distance of 80 Mm, and the average speed and lifetime are about \speed{100} and 30 minutes, respectively. \cite{podladchikova10} reported that small-scale EUV waves can be detected up to a distance of 40 Mm over a lifetime of 20 minutes, and they interpreted these small-scale EUV waves as slow-mode waves according to their average speed (\speed{14}). \cite{zhang11} also reported many small-scale EUV waves associated with EUV cyclones over the quiet Sun at a speed of \speed{35--85}, much slower than large-scale EUV waves originated from active regions. The driving source of small-scale EUV waves are found to be associated with miniature eruptions such as mini-filaments, micro-sigmoids, jets, and EUV cyclones. However, how these miniature solar eruptive activities trigger small-scale EUV waves are still unclear. In a few studies, some authors claimed that small-scale EUV waves are triggered by mini-CMEs launched by the miniature solar eruptions, which resemble their large-scale counterparts many characteristics and are interpreted as fast-mode magnetosonic waves  \citep[e.g.,][]{zheng11,zheng12a,zheng12b,zheng12c,zheng12d,zheng13a,zheng13b,zheng14}. \cite{liu14} noted that small-scale EUV waves could be divided into two categories according to their locations of the source regions. Namely, those from ephemeral regions closely resemble large-scale EUV waves can be interpreted as fast-mode magnetosonic waves, which are often associated with miniature eruptions of mini-filaments, micro-sigmoids, and coronal jets \citep{zheng11,zheng12a,zheng12b,zheng12c,zheng12d,zheng13a,zheng13b,zheng14}, and has a larger size and faster speed in the range of \speed{200--500}. Those from the quiet Sun have smaller size and slower speed of \speed{10--100} and can be interpreted as slow-mode waves \citep{podladchikova10}  or non-wave coronal reconfigurations. They are often associated with supergranular flows \citep{innes09} or coronal cyclones driven by rotating network magnetic fields \citep{zhang11,yu15}. So far, studies on small-scale EUV waves are still very scarce, questions such as their driving mechanism, physical nature, their relation to the variation of photosphere magnetic field, and their role played in the full spectrum of EUV waves are still unclear. Therefore, more detailed observational studies based on high resolution observations are needed.

In the present study, we present the observational analysis of a small-scale EUV wave that occurred on March 21, 2016 from a region of quiet Sun. It was associated with a mini-filament eruption and a {\em GOES} B1.9 micro-flare, and the magnetic field of the eruption source region showed flux emergence and cancellations. The observational results suggest that the small-scale EUV wave should be a fast-mode magnetosonic wave that was driven by the fast expansion of a coronal loop system associated with the eruption of a mini-filament. In addition, the small-scale EUV wave also caused the transverse oscillation of a remote filament. These results provide new clues for diagnosing the driving mechanism and the physical nature of the small-scale EUV wave and the property of the oscillating filament. The instruments and observations are briefly introduced in Section 2, observational results are described in Section 3, conclusions and discussions are given in the last section.

\section{Instruments and Observations}
The event was simultaneously recorded by the Solar Magnetic Activity Research Telescope \citep[SMART;][]{ueno04} and the {\em Solar Dynamics Observatory} \citep[{\em SDO};][]{pesnell12}. The SMART is a ground-based solar telescope at Hida Observatory of Kyoto University, Japan, which observes the full-disk Sun with the H$\alpha$ line. On March 21, SMART only provided H$\alpha$ line-center observations. The cadence and pixel resolution of the H$\alpha$ images are 1 minute and $0\arcsec.6$, respectively. The line-of-sight (LOS) magnetograms and EUV observations used in the present paper is taken by the Helioseismic and Magnetic Imager \citep[HMI;][]{scho12} and the Atmospheric Imaging Assembly \citep[AIA;][]{leme12} onboard the {\em SDO}, respectively. The measurement precision of the magnetograms is 10 Gauss, and the cadence and pixel resolution are 45 seconds and $0\arcsec.6$, respectively. The AIA images the full-disk Sun up to \rsun{1.3} in seven EUV and three UV-visible channels, and it has a cadence of 12 seconds and a pixel resolution of $0\arcsec.6$. The soft X-ray 1 -- 8 \AA\ flux is recorded by the {\em Geostationary Operational Environmental Satellite} ({\em GOES}). All images used in this paper are differentially rotated to the reference time of 23:55:00 UT on March 21, and the solar north is up, west to the right.

\begin{figure*}[thbp]
\epsscale{0.9}
\figurenum{1}
\plotone{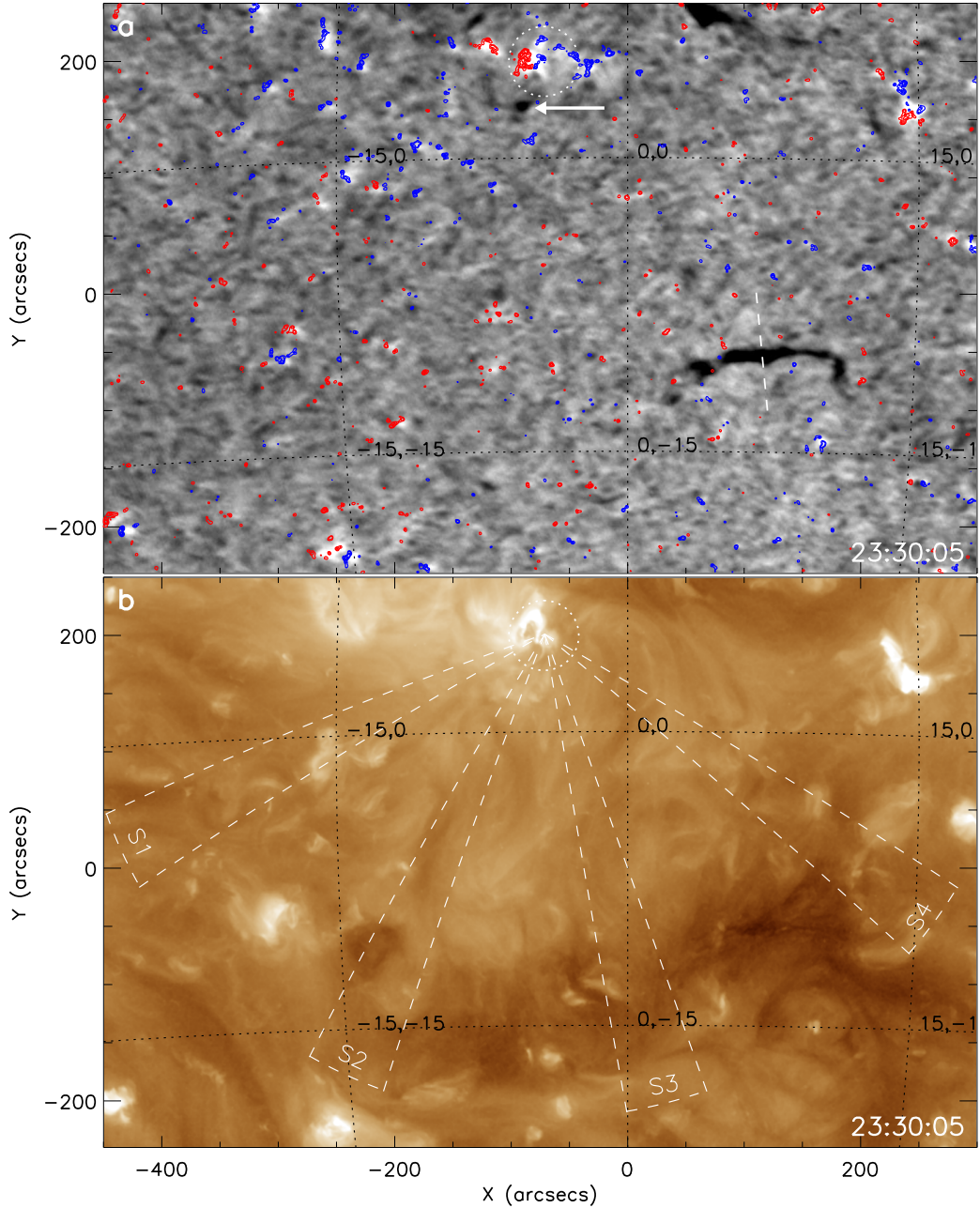}
\caption{The top and bottom panels are SMART H$\alpha$ line-center and AIA 193 \AA\ images before the eruption, respectively. The line-of-sight magnetic field at 23:40:14 UT is overlaid in panel (a), in which the red and blue contours are positive and negative magnetic fields, and the contour levels are $\pm 50$, $\pm 100$, $\pm 200$, and $\pm 300$ Gauss. The white dotted circle marks the eruption source region, and the four sectors in panel (b) show the paths for making time-distance diagrams as shown in \nfig{fig5}. The white dashed line perpendicular to the filament is used to make time-distance diagrams as shown in \nfig{fig6}. The black dotted curves in each panel are the lines of longitude and latitude. The field-of-view of each panel is $750\arcsec \times 490\arcsec$. An animation is available for this figure in the online journal.
\label{fig1}}
\end{figure*}

\section{Observational Results}
On March 21, 2016, {\em SDO} observed a mini-filament eruption in a quiet Sun region close to the solar disk center. The mini-filament eruption did not cause any CMEs in coronagraph observations, and the {\em GOES} soft X-ray 1 -- 8 \AA\ flux indicates that the eruption was accompanied by a B1.9 micro-flare, whose start and peak times are about 23:45:00 UT on March 21 and 00:05:00 UT on March 22. It is interesting that an arch-shaped EUV wave was caused by such a miniature eruption. The EUV wave can be clearly detected within a large distance over 300 Mm, and it further caused the transverse oscillation of a remote filament that located at about 240 Mm far from the eruption source region.

The source region before the eruption is shown in \nfig{fig1} with an H$\alpha$ center and an AIA 193 \AA\ images. It can be seen that the source region shows as a small bright patch in the H$\alpha$ image, which is highlighted by a white dotted circle in \nfig{fig1}. A mini-filament can be identified as a small dark patch close to the south side of the source region (see the white arrow in \nfig{fig1} (a)). In the meantime, another relatively longer filament at the south-west of the source region is also observed.  The projection distance from the filament to the eruption source region is about 240 Mm, and the filament showed obvious transverse oscillation due to the interaction of the EUV wave originated from the eruption source region. In order to show the magnetic field condition of the eruption source region and the region where the EUV wave propagates, the HMI LOS magnetic field at 23:40:14 UT is overlaid on the H$\alpha$ image (see \nfig{fig1} (a)), in which the red and blue contours represent the positive and negative magnetic fields, respectively. As one can see that the entire region of interest belongs to a quiet Sun region, and the magnetic field in the eruption source region is relatively stronger than the surrounding region and shown as a small bipolar region. In the AIA 193 \AA\ image (\nfig{fig1} (b)), the source region is observed as a small bright loop system, and the mini-filament observed in H$\alpha$ images can not be identified.

\begin{figure*}[thbp]
\epsscale{0.9}
\figurenum{2}
\plotone{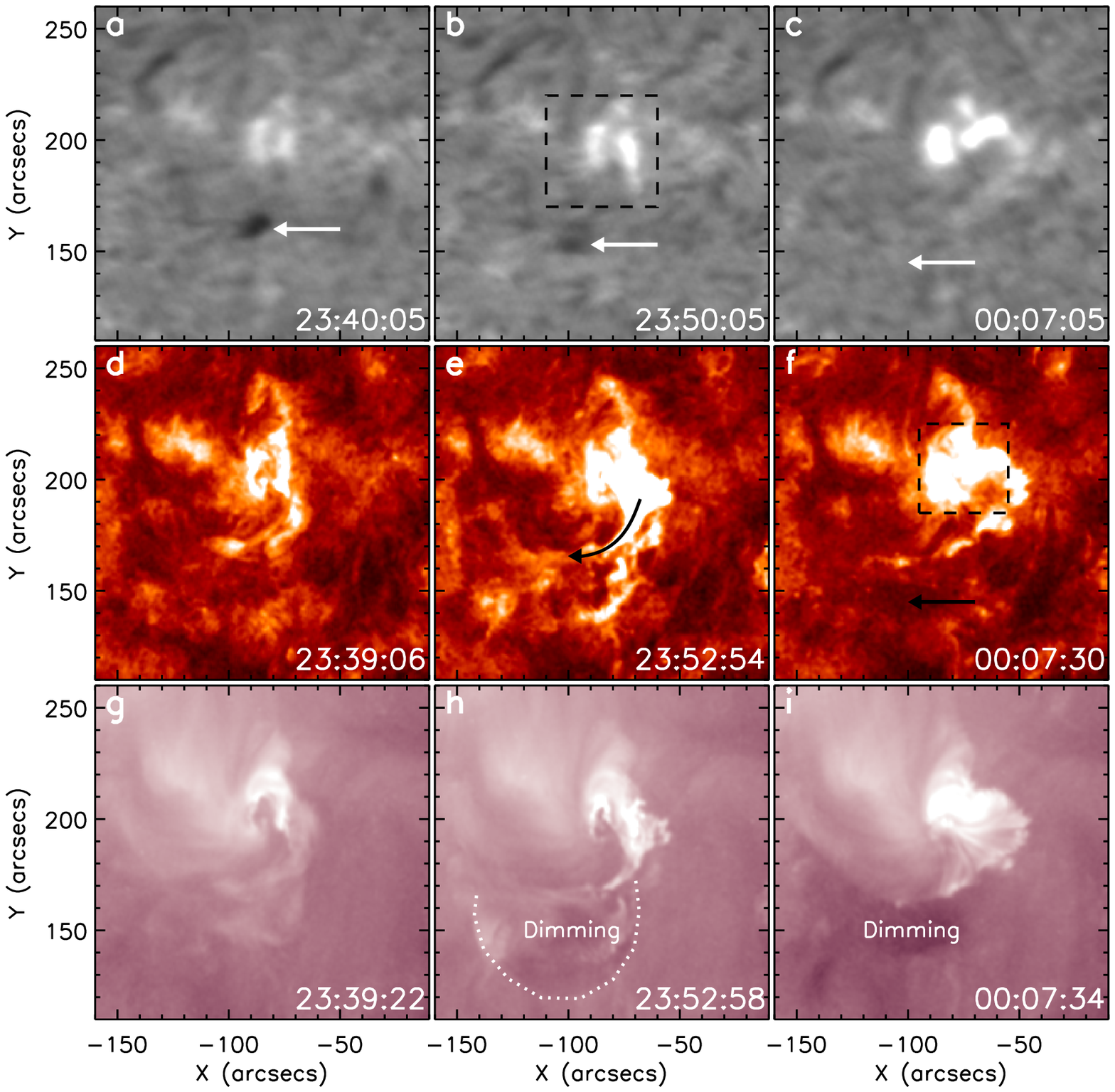}
\caption{The top row, middle, and bottom rows are the SMART H$\alpha$ line-center, AIA 304 \AA\, and AIA 211 \AA\ images, respectively. The black dashed box in panel (b) shows the field-of-view of the magnetograms shown in \nfig{fig3}. The horizontal arrows in the top row indicate the mini-filament, and the arrows in panel (e) indicates the ejecting direction of the small ejector from the source region, and the arrow in panel (f) indicates the erupting filament. The dotted curve in panel (h) marks the expanding loop. The field-of-view of each panel is $150\arcsec \times 150\arcsec$. An animation is available for this figure in the online journal.
\label{fig2}}
\end{figure*}

\nfig{fig2} shows the eruption process of the mini-filament with different wavelength observations. In the H$\alpha$ images (top row), it can be seen that the dark mini-filament first moved to the south and then erupted. At about 00:07:05 UT, the filament had totally disappeared from the H$\alpha$ images. During the eruption, two bright flare ribbons are observed in the eruption source region (see the top row of \nfig{fig2}), which indicate the magnetic reconnection process underneath the erupting filament like the physical picture described in large-scale filament eruptions \citep[e.g.,][]{lin00,lin03,shen11,shen12d}, as well as mini-filament eruption involving in coronal blowout jets \citep[e.g.,][]{shen12e}. In the AIA 304 \AA\ observations (middle row), a moving jet-like feature is observed from the eruption source region to the position of the mini-filament before the filament eruption, and the trajectory of the jet-like feature is indicated by the black arrow in \nfig{fig2} (e). It is found that the onset of the mini-filament eruption was due to the disturbance of this small jet-like feature. The interaction between the small plasma ejecta and the mini-filament occurred at about 23:52:54 UT, after that the mini-filament underwent a rising and eruption process. The erupting filament is indicated by the black arrow in \nfig{fig2} (f). By comparison the H$\alpha$ (00:07:05 UT) and AIA 304 \AA\ (00:07:30 UT) images at the same time, it is found that the erupting filament can be obviously observed in the AIA 304 \AA\ image but invisible in the H$\alpha$. The EUV observation of the eruption is shown with AIA 211 \AA\ images (bottom row). The most obvious characteristics of the eruption in EUV observations include (1) the expansion of a loop system followed by a dark dimming region, (2) the formation of a bright post-flare-loop in the source region, and (3) the generation of an EUV wave ahead of the expanding coronal loop. The expanding loop is highlighted by the dotted curve in \nfig{fig2} (h), and the bright post-flare-loop is obvious in \nfig{fig2} (i).  For more details about the eruption process, one can see the animation available in the online journal.

\begin{figure*}[thbp]
\epsscale{0.9}
\figurenum{3}
\plotone{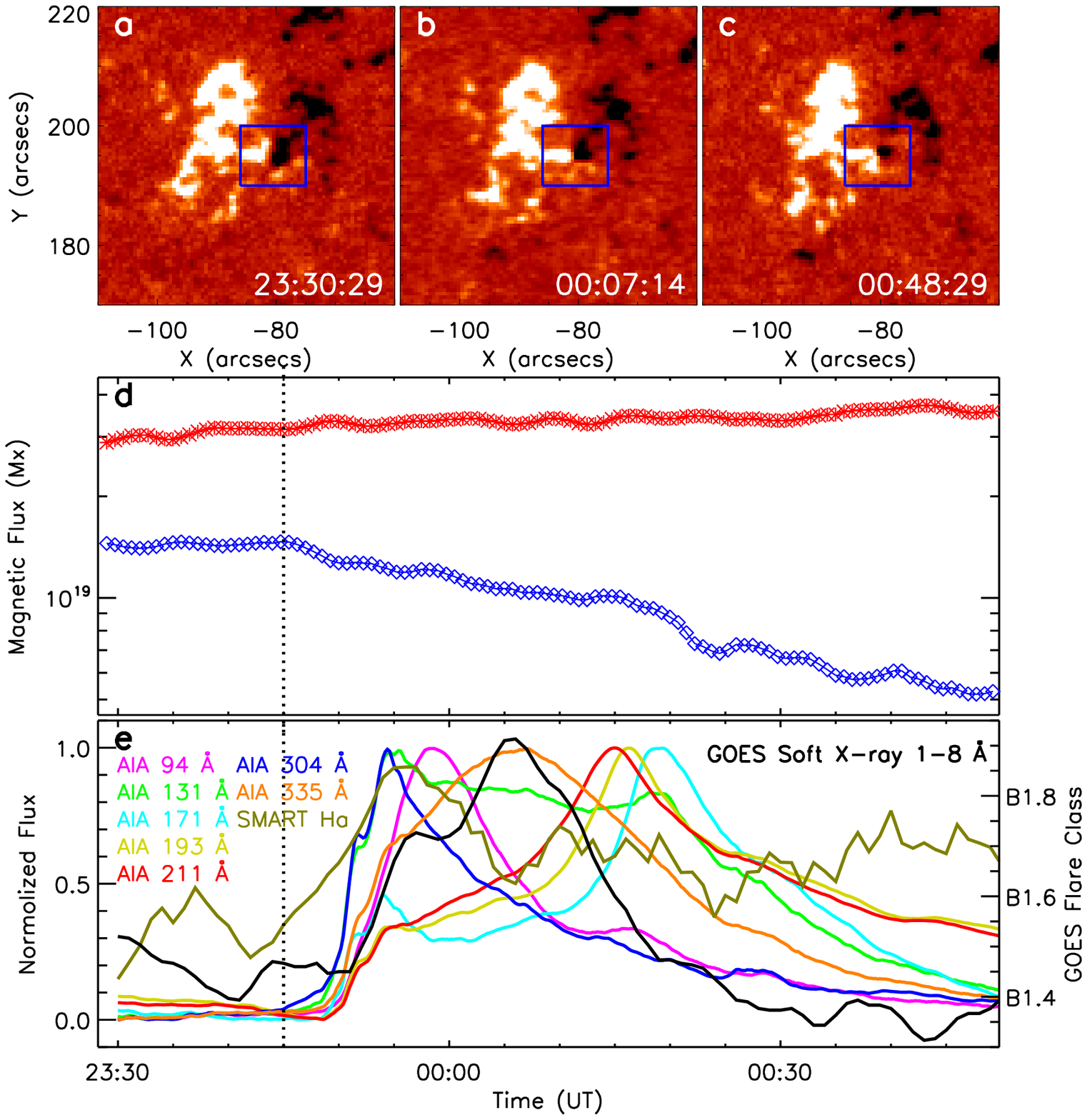}
\caption{The top row is line-of-sight magnetograms. The blue box in panels (a) -- (c) show the region used to calculate the magnetic fluxes shown in panel (d). The red asterisks and blue diamonds are the positive and negative magnetic fluxes within the box region, respectively. Panel (e) shows the lightcurves of the eruption source region measured from the seven AIA EUV channels and the {\em GOES} soft X-ray 1 -- 8 \AA\ flux (black) in the box region as shown in \nfig{fig2} (f). The vertical dashed line in panels (d) and (e) marks the start time of the eruption. The field-of-view of each panel in the top row is $50\arcsec \times 50\arcsec$.
\label{fig3}}
\end{figure*}

The evolution of the magnetic field in the eruption source region is analyzed by using the HMI LOS magnetograms, and the results are shown in the top row of \nfig{fig3}. In HMI LOS magnetograms, the white and black patches represent the positive and negative magnetic polarities, respectively. It is obvious that the area of the negative magnetic polarity became smaller during the eruption, which indicates that magnetic flux cancellation occurred due to the intruding of the positive magnetic polarity from the east direction (compare the box regions of panels (a) and (c) in \nfig{fig3}). The variations of the positive and the absolute value of the negative magnetic fluxes are plotted in \nfig{fig3} (d). It can be seen that the positive magnetic flux shows increasing trend during the entire observing time interval, which suggests the emerging process of the positive polarity. For the negative magnetic flux, it shows a sudden decrease just at around the beginning of the eruption (23:45:00 UT; see also the vertical dotted line in \nfig{fig3}). In the meantime, the positive magnetic flux shows a short time of rapid increase. This indicates that the rapid emerging of the positive magnetic polarity caused the interaction and flux cancellation between the positive and negative polarities, which triggered the ejection of the jet-like feature and therefore the eruption of the mini-filament. The lightcurves within the eruption source region (black box in \nfig{fig2} (f)) measured from AIA's different EUV channels, H$\alpha$, and the {\em GOES} soft X-ray 1 -- 8 \AA\ flux are plotted in panel (e) of \nfig{fig3}. It indicates that only the 304 \AA\ and H$\alpha$ lightcurves started to increase simultaneously with the beginning of the magnetic cancellation, the lightcurves of other AIA channels showed a few minutes delay. This indicates that the magnetic cancellation first caused the small plasma ejecta observed in the H$\alpha$ and AIA 304 \AA\ images, then the eruption of the mini-filament resulted in the delayed rising of other lightcurves of AIA channels. Considering the tight temporal relation between the start times of the magnetic cancellation and the beginning of the eruption, we can conclude that the photospheric magnetic flux cancellation resulted in the instability of the magnetic system and the initiation of the eruption.

\begin{figure*}[thbp]
\epsscale{0.9}
\figurenum{4}
\plotone{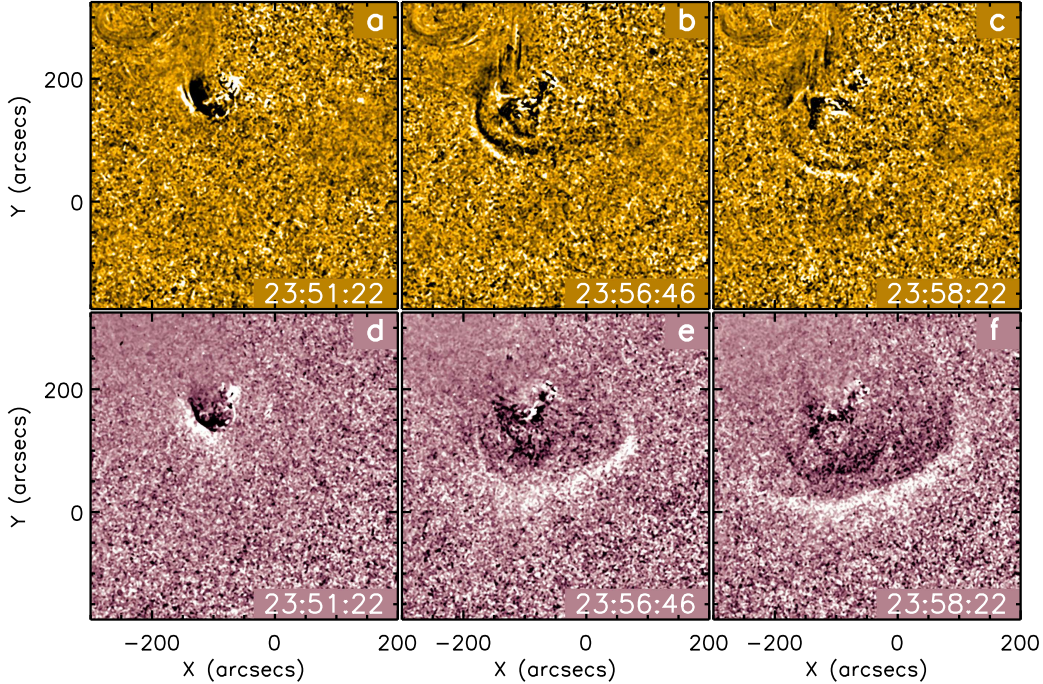}
\caption{The morphological evolution of the EUV wave. The top and bottom rows are the AIA 171 \AA\ and 211 \AA\ running ratio images, respectively. The field-of-view of each panel is $500\arcsec \times 500\arcsec$. An animation is available for this figure in the online journal.
\label{fig4}}
\end{figure*}

The evolution of the EUV wave is shown in \nfig{fig4} using the AIA 171 \AA\ and 211 \AA\ running ratio  images. Here a running ratio image is obtained by dividing the current image by the previous one in time. It is noted that the EUV wave can only be detected at the AIA 193, and 211 \AA\ observations, which suggests the narrow temperature and height range dependence of this EUV wave. Here, we only show the EUV wave with the AIA 211 \AA\ running ratio images. In the AIA 171 \AA\ running ratio images, the expansion coronal loop can be clearly observed, which moved in the southeast direction and followed by a dark dimming region. In the AIA 211 \AA\ running ratio images, the expanding coronal loop can only be clear identified during the very beginning stage of the eruption. At about 23:56:46 UT, a bright EUV wave became obviously ahead of the expansion loop during the fast expansion phase of the coronal loop, but it more clearly in the southwest direction. The first appearance time of the EUV wave delays the start times of the mini-filament eruption and the associated flare about 3 and 10 minutes, respectively. This suggests that the EUV wave was probably excited by the fast expanding coronal loop system, which represents the prototype of a CME in the low corona. Therefore, the same with many studies of large-scale EUV waves, this small-scale EUV wave is also driven by the expanding CME in the low corona rather than the flare induced pressure pulse \citep[][]{cheng12,ma11,shen13a,shen14b}. From another point of view, it is also hard to imagine that the EUV wave was driven by the pressure provided by such a micro-flare. It should be pointed out that it is hard to distinguish the separation process of the EUV wave from the expanding loop in the present case like the limb events reported by \cite{ma11} and \cite{cheng12}. This is probably caused by two reasons: one is the small-scale nature of the present event, in which the expanding loop and the preceding EUV wave have similar speeds during the beginning stage; the other is that the present event was on the solar disk, the projection effect may be important for us to distinguish the separation process.

\begin{figure*}[thbp]
\epsscale{0.9}
\figurenum{5}
\plotone{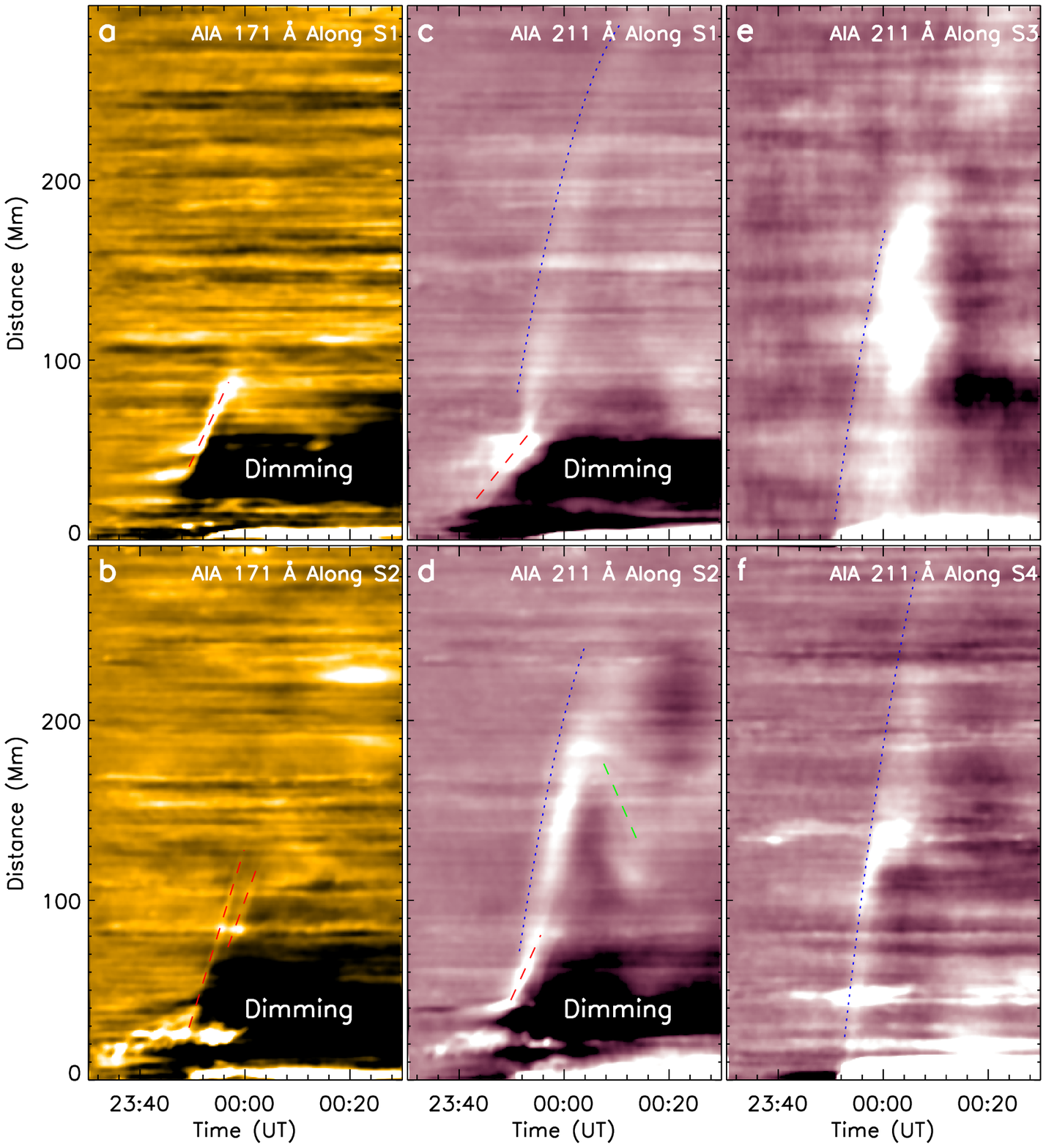}
\caption{Time-distance diagrams show the kinematics of the expanding loop system and the EUV wave  along sectors S1 -- S4 as shown in \nfig{fig1} (b). Panels (a) and (b) are AIA 171 \AA\ time-distance diagrams along S1 and S2, respectively. Panels (c) -- (f) are AIA 211 \AA\ time-distance diagrams made along S1 -- S4, respectively. The green dashed line is a linear fit to the reflected wave. The dashed red lines in the figure are the linear fit to the stripes that represent the expanding loop, while the dotted blue curve the second order fitting of a polynomial of the EUV wave.
\label{fig5}}
\end{figure*}

The kinematics of the expanding loop system and the EUV wave are analyzed using time-distance diagrams made from the AIA 171 and 211 \AA\ running ratio images along the four sectors (S1 -- S4) shown in \nfig{fig1} (b). In the AIA 171 \AA\ time-distance diagrams (\nfig{fig5} (a) and (b)), the expanding loop system shows as an inclined bright stripe, behind of which is the dark dimming region. The expansion of the loop system underwent a short acceleration and then expanding at a constant speed of about \speed{104} along S1 (\nfig{fig5} (a)). The moving speed of the expanding loop is obtained by fitting the bright stripe in the time-distance diagram with a linear function. Along S2 (\nfig{fig5} (b)), one can observe two expanding loops, and the expansion speeds of the ahead and behind loops are \speed{157 and 134}, respectively. The pair of expanding loops can also be observed in the direct imaging observations (see \nfig{fig4} (b) and (c)). In the AIA 211 \AA\ time-distance diagrams along S1 (\nfig{fig5} (c)) and S2 (\nfig{fig5} (d)), both the expanding loop ( red dashed lines) and the EUV wave (blue dotted curves) can be observed simultaneously. It is measured that the expanding speeds of the loop system along S1 and S2 are about \speed{60 and 105}, respectively. In both AIA 171 and 211 \AA\ time-distance diagrams, the dimming region firstly expanded as the expansion of the loop system, then it stopped at the distances of 58 Mm and 75 Mm along the paths of S1 and S2, respectively. 

\begin{figure*}[thbp]
\epsscale{0.9}
\figurenum{6}
\plotone{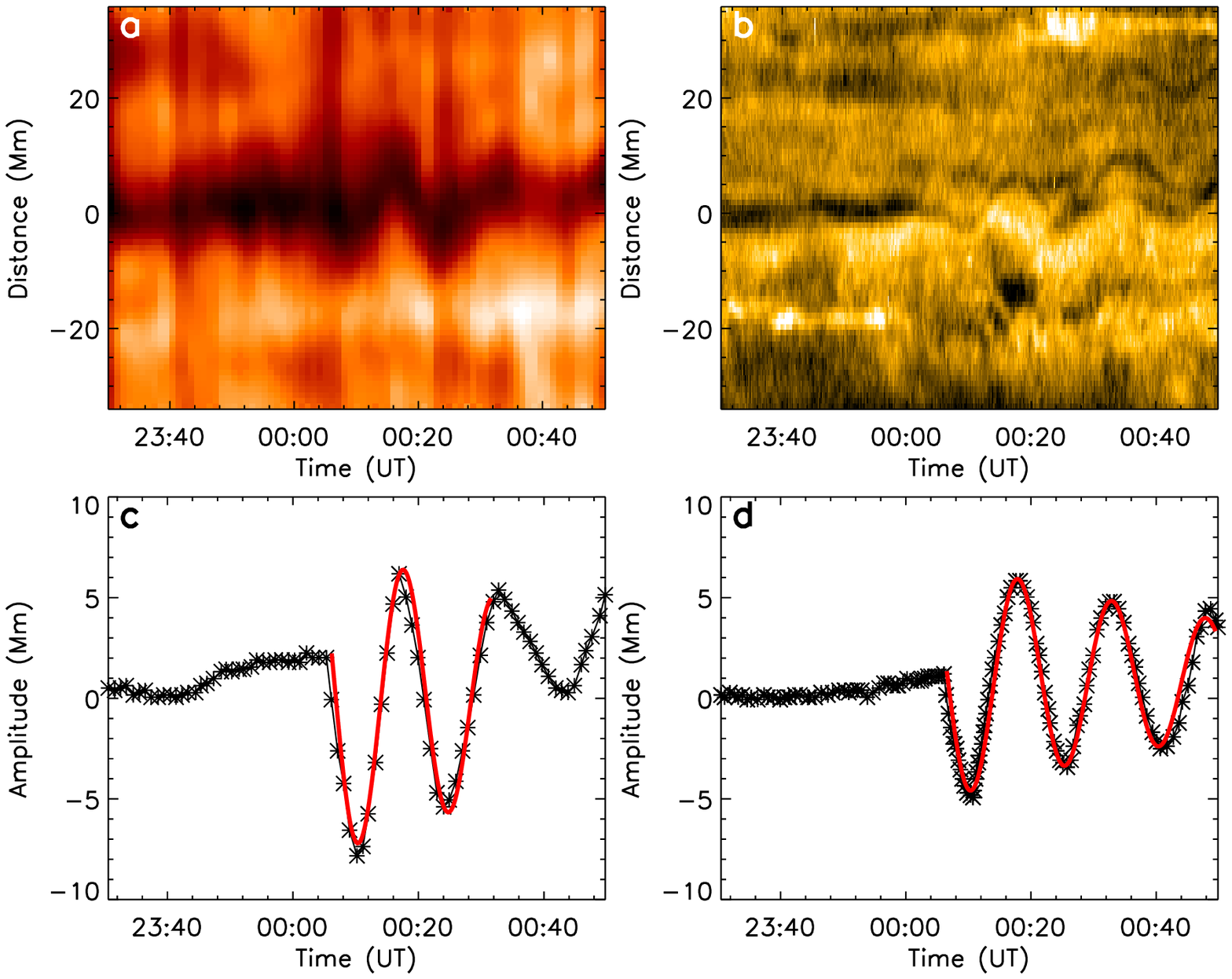}
\caption{The analysis of the oscillating filament. Panels (a) and (b) are time-distance diagrams made from SMART H$\alpha$ line-center direct and AIA 171 \AA\ running ratio images along the path perpendicular to the filament axis, respectively. Panels (c) and (d) show the oscillation trajectory of the filament determined from the H$\alpha$ and AIA 171 \AA\ time-distance diagrams, respectively. In panels (c) and (d), the measured data points are plotted as black asterisks, while the red curves are the fitting of the data points with a vibration equation in form of $F(t) = A \exp(-\frac{t}{\tau}) \sin(\omega t+\phi)$.
\label{fig6}}
\end{figure*}

For the EUV wave, it generated ahead of the fast expansion phase of the loop system at a distance of about 60 Mm from the eruption source region. The EUV wave can be detected in a distance range of about 60 Mm to over 300 Mm, and its propagation showed obvious deceleration (see panels (c) to (f) of \nfig{fig5}). Along the path of S2, the EUV wave also showed reflection due to the interaction with a small coronal magnetic structure in the path (see also \nfig{fig1} (b)). The phenomenon of reflection of EUV waves can be observed during their interaction with other magnetic structures such as active regions, coronal holes, as well as coronal bright points. Detailed analysis of this phenomenon can be found in many articles \citep[e.g.,][]{gopalswamy09,shen12a,shen12b,shen13a,li12b,kienreich13}. The blue dotted curves in \nfig{fig5} are the second order polynomial fitting to the propagating EUV wavefront, based on which we obtain that the accelerations of the EUV wave are about \accel{-90, -120, -123, and -133} along the paths of S1 to S4, respectively. The acceleration of this EUV wave is lower than the ``typical'' value of \accel{-200} for large-scale EUV waves \citep[e.g.,][]{shen12b,warmuth15}. The average propagation speeds along the four paths are also calculated based on linear fit to the data points, and the results are \speed{182, 231, 282, and 317}, respectively. Obviously, the propagation speed of the EUV wave is faster than that of the expansion of the loop system and higher than the sound speed in the quiet Sun corona (\speed{150 -- 210} for 1 -- 2 MK coronal temperature). The lower acceleration and speed of the present EUV wave relative to the large-scale EUV waves are probably due to the less energetic nature of the associated miniature eruption. However, the deceleration, the reflection, and the fast speed of the EUV wave together suggest that this small-scale EUV wave should be a fast-mode magnetosonic wave which was driven by the expanding coronal loop system.  The speed of the reflected wave is about \speed{120}, which is relatively slower than the incident wave's speed (\speed{180}) just before the interaction. Although the speed of the reflected wave is smaller than the incident wave and the sound speed in quiet Sun, we can still consider it as a wave due to its similar speed with the incident wave. Generally speaking, the wave will partially loss its energy during the interaction, and the propagation direction of the reflected wave may not along the path that we used to measure the wave speed. These reasons can cause the slower speed of the reflected wave than the incident wave. Such a phenomenon is also reported in many previous studies \citep[e.g.,][]{gopalswamy09,shen12a,kienreich13}.

It is interesting that the propagation of the EUV wave further caused the transverse oscillation of a remote  filament on the path. The filament located at a distance of about 240 Mm from the eruption source region in  the southwest direction, which can be observed clearly in the H$\alpha$ and EUV images (see \nfig{fig1}). The transverse oscillation of this remote filament started about 10 minutes after the generation of the EUV wave. Since the distance between the filament and the generation position of the EUV wave is about 180 Mm, the required propagation speed of the EUV wave to trigger the oscillation of the filament should be \speed{300}, which is well in agreement with the measured speed (\speed{317}) of the EUV wave along S4 that  passes through the location of the filament (see also \nfig{fig1} (b)). The filament oscillation is analyzed using time-distance diagrams made from H$\alpha$ and AIA 171 \AA\ observations along a path across the filament main axis, and the results are displayed in \nfig{fig6}. At about 00:05:15 UT on March 22, the EUV wave interacted with the filament. The filament was immediately pushed to the south direction and then started to oscillate for about three cycles. The oscillating filament can well be identified in the time-distance diagrams as shown in panels (a) and (b) of \nfig{fig6}. The trajectories of the oscillating filament are traced from the H$\alpha$ and AIA 171 \AA\ time-distance diagrams and plotted in panels (c) and (d) of \nfig{fig6}, respectively. By fitting the data points (black asterisk) with a damped vibration equation in form of $F(t) = A \exp(-\frac{t}{\tau}) \sin(\omega t+\phi)$, the oscillation parameters of the filament can be obtained. The fitting results indicate that the amplitudes ($A$), periods ($T$), and the damping times ($\tau$) of the oscillating filament obtained from the H$\alpha$ (AIA 171 \AA\ ) data are 8 (6) Mm, 14.4 (15) minutes, and 60 (60) minutes, respectively. The oscillation parameters of the filament are consistent with those reported in previous studies \citep[][]{shen14a,shen14b,tripathi09}.

With the measured oscillation parameters of the filament, we can estimate the radial component of the filament's magnetic field with the filament seismology technique proposed by \cite{hyde66}. The mathematic relation between the radial magnetic field and the oscillation period and damping time can be written as $B_{\rm r}^{2} = \pi \rho \, r_{\rm 0}^{2} \, [4 \, \pi ^{2} \, (\frac{1}{T})^{2} + (\frac{1}{\tau})^{2}]$, where $B_{\rm r}$ is the radial component of the filament magnetic field, $\rho$ is the density of the filament mass,  $r_{\rm 0}$ is the scale height of the filament, $\rm T$ is the oscillation period, and $\tau$ is the damping time. This equation can be wrote as $B_{\rm r}^{2} = 4.8 \times 10^{-12} \, r_{0}^{2} \, [(\frac{1}{T})^{2} + 0.025 \, (\frac{1}{\tau})^{2}]$, if we assume $\rho = 4 \times 10^{-14} \, {\rm g \ cm^{-3}}$, i.e., $n_{\rm e} = 2 \times 10^{10} \, {\rm cm^{-3}}$. In addition, the same with \cite{hyde66}, we can set the value $r_{\rm 0} = 3 \times 10^{9}$ cm. With the measured period and damping time of the filament, it is estimated that the radial component of the filament magnetic field is about 7 Gauss, which is consistent with the values obtained by other methods \citep[e.g.,][]{warw65,lee65}.

\section{Conclusions and Discussions}
Using high temporal and spatial resolution observations taken by the SMART and {\em SDO}, we present the observational analysis of a small-scale EUV wave that was in association with a mini-filament eruption in the quiet Sun. This eruption accompanied by a {\em GOES} soft X-ray B1.9 micro-flare and without CME association. The eruption source region is identified as a small bipolar magnetic region, in which magnetic flux emergence and cancellation are detected and they are thought to be the cause of the instability and eruption of the magnetic system. The photospheric magnetic flux cancellation first resulted in the ejection of a small plasma ejecta in the source region, which impacted on a nearby mini-filament on the south and further caused the filament's  rising and eruption. During this period, an expanding loop system followed by a dark dimming region is observed in the southeast direction, which underwent a short acceleration and then a fast expansion phases. It is interesting that an EUV wave was generated ahead of the expanding loop system during the fast expansion phase, and it further resulted in the transverse oscillation of a remote filament that located at a distance of about 240 Mm from the eruption source region. 

It is found that both the expanding loop system and the following dimming region stopped at a finite distance less than 100 Mm from the center of the eruption source region, and the expanding speed of the loop system is about \speed{104 -- 157}. The EUV wave generated ahead of the expanding loop at a distance of about 60 Mm and propagated to a large distance over 300 Mm from the center of the eruption source region, and its propagation showed obvious deceleration and reflection behaviors. Measurements results indicate that the acceleration and average speed of the EUV wave are about \accel{-90 -- -133} and   \speed{182 -- 317}, respectively. It is found that the acceleration is lower than the ``typical'' value (\accel{-200}) obtained in statistical studies of large-scale EUV waves \citep{warmuth15}. In the meantime, the propagation speed is also lower that the statistical average speed (\speed{644}) of large-scale EUV waves \citep{nitta13}. The lower acceleration and speed of the present EUV wave relative to their large-scale counterparts are probably due to the less energetic nature of this miniature eruption. However, the present small-scale EUV wave has similar physical properties as those observed in large-scale EUV wave events, such as the phenomena of deceleration, reflection, and the oscillation of the remote filament. In addition, although the wave speed is lower than their large-scale counterparts, the propagation speed of the present EUV wave is still faster than the expanding loop system and  the sound speed in the quiet Sun corona. Therefore, we propose that the EUV wave observed in the present event should be a fast-mode magnetosonic wave driven by the expanding coronal loop. 

It is found that the first appearance time of the EUV wave delays the start time of the associated flare about  10 minutes. Therefore, it is unlikely that this small-scale EUV wave was excited by the flare pressure pulse due to the lower flare class (B1.9) and the large time interval between the two phenomena. For large-scale EUV waves, previous studies suggested that some of them are associated with small flares, but the probability of a given small flare having an associated EUV wave is very low \citep{cliver05}. In the statistical study presented by \cite{chen06}, he found that only those flares accompanied by CMEs can generate EUV waves, all energetic large flares in their sample without CMEs have no EUV wave associations. The present study highly support the scenario that EUV waves are driven by CMEs \citep{chen06,nitta13,liu14}, although there was no CME in the present case. For small-scale EUV waves, they often associate with micro-flares and miniature solar eruptions that have no enough energy to launch large-scale CMEs. For example, \cite{zheng12b} found that a fast-mode EUV wave is associated with a failed filament eruption that has no CME association. In other studies, it is observed that small-scale EUV waves can be excited by newly formed expanding loops through tether-cutting mechanism \citep[e.g.,][]{zheng12d,zheng13b}. In the present event, the temporal and spatial relationship between the wave and the expanding loop system provide evidence that the EUV wave was driven by the expanding loop system. We emphasize that the expanding loop observed in the present event and the newly formed loops in \cite{zheng12d} and \cite{zheng13b} can be considered as the CME prototype in the low corona. Although miniature solar eruptions involving expanding loops can not launch successful CMEs that can be observed by coronagraphs, they resemble many characteristics with large-scale successful CMEs during the initiation eruption stage.

Our analysis results indicate that the transverse oscillation of the remote filament should be launched by the interaction of the EUV wave. Measurement results based on the H$\alpha$ (AIA 171 \AA\ ) observatoins suggest that the amplitude, period, and damping time of the oscillation filament are 8 (6) Mm, 14.4 (15) minutes, and 60 (60) minutes, respectively. These parameters of the oscillation filament are consistent with the values reported in previous studies on filament oscillations caused by large-scale EUV waves \citep[][]{shen14a,shen14b,tripathi09}. With the method of filament seismology, it is estimated that the radial component of the filament magnetic field is about 7 Gauss. In previous studies, transverse filament oscillations are thought to be driven by chromosphere Moreton waves \citep[e.g.,][]{moreton60,hyde66}, while longitude oscillations are caused by near-by flares, jets, and other activities \citep[e.g.,][]{jing03,vrsn07,li12a,bi14,zhang12,zhang17}. With coronal EUV observations, it is found that energetic large-scale coronal EUV wave can also trigger filament oscillations \citep[e.g.,][]{eto02,okam04,shen12b}. Recently, \cite{shen14a} found that a chain of oscillating filaments are caused by a very weak EUV wave that was associated with an energetic {\em GOES} X2.1 flare. The authors also proposed that the interaction angle between the incoming EUV wave and the filament main axis is important to trigger what types of filament oscillation. Namely, if the propagation direction of an EUV wave is perpendicular (parallel) to the filament axis, transverse (longitudinal) oscillation of the filament can be expected \citep{shen14b,pant16}. In the present event, the EUV wave interacted with the filament perpendicularly and resulted the transverse oscillation of the filament, in agreement with the model proposed by \cite{shen14b}. 

In summary, we conclude that the present small-scale EUV wave should be a fast-mode magnetosonic wave that was driven by the expanding coronal loop. It showed similar characteristics with large-scale EUV waves such as deceleration and reflection effect during the propagation, but the speed and acceleration of the present small-scale EUV wave are much slower than the those of large-scale EUV waves. The present event also suggests that filament oscillations can not only be triggered by large-scale energetic EUV and Moreton waves, but also small-scale EUV waves associated with less energetic miniature solar eruptions. So far, detailed studies on small-scale EUV wave are still very scarce, although their occurrence rate is very high. In the future, more observational and statistical studies are desirable for comparing the similarities and differences between small- and large-scale EUV waves about their driving source, excitation mechanism, and physical properties. In addition, since small-scale EUV waves often associate with micro-flares in the quiet Sun, resembling the size distribution of flares \citep[e.g.,][]{hudson91,aschwanden07}, small-scale EUV waves are probably also important in the full spectrum of EUV waves of hierarchic sizes.

\acknowledgments We thank the excellent observations provided by the SMART and the {\em SDO}, and the anonymous referee for his/her suggestions and comments that largely improve the quality of the present paper. This work is supported by the Natural Science Foundation of China (11403097,11633008,11773068), the Yunnan Science Foundation (2015FB191,2017FB006), the Specialized Research Fund for State Key Laboratories, and the Youth Innovation Promotion Association (2014047) of Chinese Academy of Sciences. Z. Qu is supported by the research fund of Sichuan University of Science and Engineering (Grant Nos. 2015RC43).


\begin{thebibliography}{}
\bibitem[Aschwanden(2007)]{aschwanden07}
Aschwanden, M.~J.\ 2007, Advances in Space Research, 39, 1867 
\bibitem[Attrill(2010)]{attr10}
Attrill, G. D. R. 2010, \apj, 718, 494
\bibitem[Attrill et al.(2009)]{attr09}
Attrill, G. D. R., Engell, A. J., Wills-Davey, M. J., Grigis, P., \& Testa, P. 2009, \apj, 704, 1296
\bibitem[Attrill et al.(2007)]{attr07}
Attrill, G. D. R., Harra, L. K., van Driel-Gesztelyi, L., \& D\'{e}moulin, P. 2007, \apj, 656, L101
\bibitem[Bi et al. (2014)]{bi14}
Bi, Y., Jiang, Y., Yang, J., et al. 2014, \apj, 790, 100
\bibitem[Chandra et al. (2016)]{chandra16}
Chandra, R., Chen, P. F., Fulara, A., Srivastava, A. K., \& Uddin, W. 2016, \apj, 822, 106
\bibitem[Chen et al.(2006)]{chen06}
Chen, P. F.\ 2006, \apjl, 641, 153
\bibitem[Chen et al.(2005)]{chen05}
Chen, P. F., Fang, C., \& Shibata, K.\ 2005, \apj, 622, 1202
\bibitem[Chen et al. (2016)]{chen16}
Chen, P. F., Fang, C., Chandra, R., \& Srivastava, A. K. 2016, \solphys, 291, 3195
\bibitem[Chen et al.(2002)]{chen02}
Chen, P. F., Wu, S. T., Shibata, K., \& Fang, C.\ 2002, \apjl, 572, 99
\bibitem[Chen \& Wu(2011)]{chen11}
Chen, P. F., \& Wu, Y.\ 2011, \apjl, 732, 20
\bibitem[Cheng et al.(2012)]{cheng12}
Cheng, X., Zhang, J., Olmedo, O., et al.\ 2012, \apjl, 745, 5
\bibitem[Cliver et al. (2005)]{cliver05}
Cliver, E. W., Laurenza, M., Storini, M., \& Thompson, B. J. 2005, \apj, 631, 604
\bibitem[Cohen et al.(2009)]{cohe09}
Cohen, O., Attrill, G. D. R., Manchester IV, W. B., \& Wills-Davey, M. J. 2009, \apj, 705, 587
\bibitem[Dai et al.(2010)]{dai10}
Dai, Y., Auch\`{e}re, F., Vial, J.-C., Tang, Y. H., \& Zong, W. G. 2010, \apj, 708, 919
\bibitem[Delann\'{e}e(2000)]{dela00}
Delann\'{e}e, C. 2000, \apj, 545, 512
\bibitem[Delann\'{e}e \& Aulanier(1999)]{dela99}
Delann\'{e}e, C., \& Aulanier, G. 1999, \solphys, 190, 107
\bibitem[Delann\'{e}e et al.(2007)]{dela07}
Delann\'{e}e, C., Hochedez, J.-F., \& Aulanier, G. 2007, \aap, 465, 603
\bibitem[Delann\'{e}e et al.(2008)]{dela08}
Delann\'{e}e, C., T\"{o}r\"{o}k, T., Aulanier, G., \& Hochedez, J.-F. 2008, \solphys, 247, 123
\bibitem[Downs et al.(2011)]{down11}
Downs, C., Roussev, I. I., van der Holst, B., et al. 2011, \apj, 728, 2
\bibitem[Eto et al.(2002)]{eto02}
Eto, S., Isobe, H., Narukage, N., et al. 2002, \pasj, 54, 481
\bibitem[Foley et al.(2003)]{fole03}
Foley, C. R., Harra, L. K., Matthews, S. A., Culhane, J. L., \& Kitai, R. 2003, \aap, 399, 749
\bibitem[Gallagher \& Long (2011)]{gallagher11}
Gallagher, P. T., \& Long, D. M. 2011, \ssr, 158, 365
\bibitem[Goddard et al. (2016)]{goddard16}
Goddard, C. R., Nistic\'{o}, G., Nakariakov, V. M., Zimovets, I. V., \& White, S. M. 2016, \aap, 594, A96
\bibitem[Gopalswamy et al.(2009)]{gopalswamy09}
Gopalswamy, N., Yashiro, S., Temmer, M., et al.\ 2009, \apjl, 691, L123 
\bibitem[Guo et al.(2015)]{guo15}
Guo, Y., Ding, M. D., Chen, P. F.\ 2015, \apjs, 219, 36
\bibitem[Harra \& Sterling(2003)]{harr03}
Harra, L. K., \& Sterling, A. C. 2003, \apj, 587, 429
\bibitem[Hudson(1991)]{hudson91}
Hudson, H.~S.\ 1991, \solphys, 133, 357 
\bibitem[Hyder(1966)]{hyde66}
Hyder, C. L. 1966, Z. Astrophys. 63, 78
\bibitem[Innes et al. (2009)]{innes09}
Innes, D. E., Genetelli, A., Attie, R., \& Potts, H. E. 2009, \aap, 495, 319
\bibitem[Jing et al.(2003)]{jing03}
Jing, J., Lee, J., Spirock, T. J., et al. 2003, \apjl, 584, L103
\bibitem[Khan \& Aurass (2002)]{khan02}
Khan, J. I., \& Aurass, H. 2002, \aap, 383, 1018
\bibitem[Kienreich et al.(2013)]{kienreich13}
Kienreich, I.~W., Muhr, N., Veronig, A.~M., et al.\ 2013, \solphys, 286, 201 
\bibitem[Kwon et al.(2013)]{kwon13}
Kwon, R.-Y., Ofman, L., Olmedo, O., et al. 2013, \apj, 766, 55
\bibitem[Kumar et al. (2013)]{kumar13}
Kumar, P., Cho, K.-S., Chen, P. F., Bong, S.-C., \& Park, S. -H. 2013, \solphys, 282, 523
\bibitem[Kumar \& Innes (2015)]{kumar15}
Kumar, P., \& Innes, D. E. 2015, \apjl, 803, L23
\bibitem[Kumar et al. (2017)]{kumar17}
Kumar, P., Nakariakov, V. M., \& Cho, K. -S. 2017, \apj, in press
\bibitem[Lemen et al.(2012)]{leme12}
Lemen, J. R., Title, A. M., Akin, D. J., et al. \solphys, 275, 17
\bibitem[Lee et al.(1965)]{lee65}
Lee, R. H., Rust, D. M., \& Zirin, H. 1965, Appl. Opt. 4, 1081
\bibitem[Li \& Zhang(2012a)]{li12a}
Li, T., \& Zhang, J. 2012a, \apjl, 760, L10
\bibitem[Li et al.(2012b)]{li12b}
Li, T., Zhang, J., Yang, S., \& Liu, W.\ 2012b, \apjl, 746, 13
\bibitem[Li et al.(2012c)]{li12c}
Li, T., Zhang, J., Yang, S., \& Liu, W.\ 2012c, RAA, 12, 104
\bibitem[Lin \& Forbes(2000)]{lin00}
Lin, J., \& Forbes, T.~G.\ 2000, \jgr, 105, 2375 
\bibitem[Lin et al.(2003)]{lin03}
Lin, J., Soon, W., \& Baliunas, S.~L.\ 2003, \nar, 47, 53 
\bibitem[Liu et al.(2010)]{liu10}
Liu, W., Nitta, N. V., Schrijver, C. J., Title, A. M., Tarbell, T. D.\ 2010, \apjl, 723, 53
\bibitem[Liu et al.(2011)]{liu11}
Liu, W., Title, A. M., Zhao, J., et al.\ 2011, \apjl, 736, 13
\bibitem[Liu et al.(2012)]{liu12}
Liu, W., Ofman, L., Nitta, N. V., et al.\ 2012, \apj, 753, 52
\bibitem[Liu \& Ofman(2014)]{liu14}
Liu, W., \& Ofman, L.\ 2014, \solphys, 289, 3233
\bibitem[Long et al.(2011a)]{long11a}
Long, D. M., DeLuca, E. E., \& Gallagher, P. T. 2011a, \apjl, 741, L21
\bibitem[Long et al.(2011b)]{long11b}
Long, D. M., Gallagher, P. T., McAteer, R. T. J., \& Bloomfield, D. S. 2011b, \aap, 531, A42
\bibitem[Long et al. (2017)]{long17}
Long, D. M., Bloomfield, D. S., Chen, P. F., et al. 2017, \solphys, 292, 7 
\bibitem[Ma et al.(2010)]{ma10}
Ma, S., Attrill, G. D. R., Golub, L., Lin, J.\ 2010, \apj, 722, 289
\bibitem[Ma et al.(2011)]{ma11}
Ma, S., Raymond, J.~C., Golub, L., et al.\ 2011, \apj, 738, 160
\bibitem[Ma, et al.(2009)]{ma09}
Ma, S., Wills-Davey, M. J., Lin, J., et al.\ 2009, \apj, 707, 503
\bibitem[Moreton(1960)]{moreton60}
Moreton, G. E.\ 1960, \aj, 65, 494
\bibitem[Moreton(1964)]{moreton64}
Moreton, G. E.\ 1964, \aj, 69, 164
\bibitem[Moses et al.(1997)]{moses97}
Moses, D., Clette, F., Delaboudini{\`e}re, J.-P., et al.\ 1997, \solphys, 175, 571 
\bibitem[Muhr et al.(2014)]{muhr14}
Muhr, N., Veronig, A. M., Kienreich, I. W., et al.\ 2014, \solphys, 289, 4563
\bibitem[Muhr et al.(2010)]{muhr10}
Muhr, N., Vr\v{s}nak, B., Temmer, M., Veronig, A. M., \& Magdaleni\'{c}, J. 2010, \apj, 708, 1639
\bibitem[Nistic\'{o} et al. (2014)]{nistico14}
Nistic\'{o}, G., Pascoe, D. J., \& Nakariakov, V. M. 2014, \aap, 569, A12
\bibitem[Nitta et al.(2013)]{nitta13}
Nitta, N. V., Schrijver, C. J., Title, A. M., Liu, W.\ 2013, \apj, 776, 58
\bibitem[Ofman et al. (2011)]{ofman11}
Ofman, L., Liu, W., Title, A., \& Aschwanden, M. 2011, \apjl, 740, L33
\bibitem[Okamoto et al.(2004)]{okam04}
Okamoto, T. J., Nakai, H., Keiyama, A., et al. 2004, \apj, 608, 1124
\bibitem[Pant et al. (2016)]{pant16}
Pant, V., Mazumder, R., Yuan, D., et al. 2016, \solphys, 291, 3303
\bibitem[Pascoe et al. (2013)]{pascoe13}
Pascoe, D. J., Nakariakov, V. M., \& Kupriyanova, E. G. 2013, \aap, 560, A97
\bibitem[Patsourakos \& Vourlidas(2009)]{pats09a}
Patsourakos, S., \& Vourlidas, A. 2009, \apjl, 700, L182
\bibitem[Patsourakos \& Vourlidas (2012)]{patsourakos12}
Patsourkos, S., \& Vourlidas, A. 2012, \solphys, 281, 187
\bibitem[Pesnell et al. (2012)]{pesnell12}
Pesnell, W. D., Thompson, B. J., \& Chamberlin, P. C. 2012, \solphys, 275, 3
\bibitem[Podladchikova et al. (2010)]{podladchikova10}
Podladchikova, O., Vourlidas, A., Van der Linden, R. A. M., W\"{u}lser, J.-P., \& Patsourakos, S. 2010, \apj, 709, 369
\bibitem[Schou et al.(2012)]{scho12}
Schou, J., Borrero, J. M., Norton, A. A., et al., \solphys, 275, 327
\bibitem[Shen et al.(2011)]{shen11}
Shen, Y.-D., Liu, Y., \& Liu, R.\ 2011, Research in Astronomy and Astrophysics, 11, 594 
\bibitem[Shen \& Liu (2012a)]{shen12a}
Shen, Y., \& Liu, Y.\ 2012a, \apjl, 752, 23
\bibitem[Shen \& Liu (2012b)]{shen12b}
Shen, Y., \& Liu, Y.\ 2012b, \apj, 754, 7
\bibitem[Shen \& Liu(2012c)]{shen12c}
Shen, Y., \& Liu, Y.\ 2012c, \apj, 753, 53
\bibitem[Shen et al.(2012a)]{shen12d}
Shen, Y., Liu, Y., \& Su, J.\ 2012a, \apj, 750, 12 
\bibitem[Shen et al.(2012b)]{shen12e}
Shen, Y., Liu, Y., Su, J., \& Deng, Y.\ 2012b, \apj, 745, 164 
\bibitem[Shen et al. (2013a)]{shen13a}
Shen, Y., Liu, Y., Su, J., et al.\ 2013a, \apjl, 773, 33
\bibitem[Shen et al. (2013b)]{shen13b}
Shen, Y., Liu, Y., Su, J., et al.\ 2013b, \solphys, 288, 585
\bibitem[Shen et al. (2014a)]{shen14a}
Shen, Y., Ichimoto, K., Ishii, T., et al.\ 2014a, \apj, 786, 151
\bibitem[Shen et al. (2014b)]{shen14b}
Shen, Y., Liu, Y. D., Chen, P. F., \& Ichimoto, K.\ 2014b, \apj, 795, 130
\bibitem[Thompson \& Myers (2009)]{thompson09}
Thompson, B. J. \& Myers, D. C.\ 2009, \apjs, 183, 225
\bibitem[Thompson et al.(1998)]{thompson98}
Thompson, B. J., Plunkett, S. P., Gurman, J. B., et al.\ 1998, \jgr, 25, 2465
\bibitem[Tripathi et al. (2009)]{tripathi09}
Tripathi, D., Isobe, H., \& Jain, R. 2009, \ssr, 149, 283
\bibitem[Ueno et al.(2004)]{ueno04}
Ueno, S., Nagata, S., Kitai, R., \& Kurokawa, H. 2004, in ASP Conf. Ser. 325, 319
\bibitem[Uchida (1968)]{uchida68}
Uchida, Y., 1968, \solphys, 4, 30
\bibitem[van Driel-Gesztelyi et al.(2008)]{van08}
van Driel-Gesztelyi, L., Attrill, G. D. R., D\'{e}moulin, P., Mandrini, C. H., Harra, L. K. 2008, ann. Geophys. 26, 3077
\bibitem[Vanninathan et al.(2015)]{vanninathan15}
Vanninathan, K., Veronig, A. M., Dissauer, K., et al.\ 2015, \apj, 812, 173
\bibitem[Veronig et al.(2010)]{vero10}
Veronig, A. M., Muhr, N., Kienreich, I. W., Temmer, M., \& Vr\v{s}nak, B. 2010, \apjl, 716, L57
\bibitem[Vr\v{s}nak et al.(2016)]{vrsnak16}
Vr\v{s}nak, B., \v{Z}ic, T., Luli\'{c}, S., Temmer, M., Veronig, A. M.\ 2016, \solphys, 291, 89
\bibitem[Vr\v{s}nak et al. (2005)]{vrsnak05}
Vr\v{s}nak, B., Magdaleni\'{c}, J., Temmer, M., et al. 2005, \apjl, 625, L67
\bibitem[Vr\v{s}nak et al.(2007)]{vrsn07}
Vr\v{s}nak, B., Veronig, A. M., Thalmann, J. K., \& \v{Z}ic, T. 2007, \aap, 471, 295
\bibitem[Vr\v{s}nak et al. (2002)]{vrsnak02}
Vr\v{s}nak, B., Warmuth, A., Braj\v{s}a, R., \& Hanslmeier, A. 2002, \aap, 394, 299
\bibitem[Wang et al.(2015)]{wang15}
Wang, H., Liu, S., Gong, J., Wu, N., \& Lin, J.\ 2015, \apj, 114
\bibitem[Wang et al.(2009)]{wang09}
Wang, H., Shen, C., \& Lin, J.\ 2009, \apj, 700, 1716
\bibitem[Warmuth (2015)]{warmuth15}
Warmuth, A.\ 2015, LRSP, 12, 3
\bibitem[Warmuth et al.(2004a)]{warmuth04a}
Warmuth, A., Vr\v{s}nak, B., Magdaleni\'{c}, J., et al.\ 2004a, \aap, 418, 1101
\bibitem[Warmuth et al.(2004b)]{warmuth04b}
Warmuth, A., Vr\v{s}nak, B., Magdaleni\'{c}, J., et al.\ 2004b, \aap, 418, 1117
\bibitem[Warwick \& Hyder(1965)]{warw65}
Warwick, J. W., \& Hyder, C. L. 1965, \apj, 141, 1362
\bibitem[Wills-Davey(2006)]{will06}
Wills-Davey, M. J. 2006, \apj, 645, 757
\bibitem[Xue et al.(2013)]{xue13}
Xue, Z. K., Qu, Z. Q., Yan, X. L., Zhao, L., \& Ma, L.\ 2013, \aap, 556, 152
\bibitem[Yang et al.(2013)]{yang13}
Yang, L., Zhang, J., Liu, W., Liu, W., \& Shen, Y.\ 2013, \apj, 775, 39
\bibitem[Yu et al. (2015)]{yu15}
Yu, X. T., Zhang, J., Li, T., \& Yang, S. H. 2015, RAA, 15, 1525
\bibitem[Yuan et al. (2013)]{yuan13}
Yuan, D., Shen, Y., Liu, Y., et al. 2013, \aap, 554, A144
\bibitem[Yuan et al. (2016)]{yuan16}
Yuan, D., Li, B., \& Walsh, R. W. 2016, \apj, 828, 17
\bibitem[Zhang \& Liu (2011)]{zhang11}
Zhang, J. \& Liu, Y. 2011, \apjl, 741, L7
\bibitem[Zheng et al. (2011)]{zheng11}
Zheng, R., Jiang, Y., Hong, J., et al.\ 2011, \apjl, 739, 39
\bibitem[Zheng et al. (2012a)]{zheng12a}
Zheng, R., Jiang, Y., Yang, J., et al.\ 2012a, \apj, 747, 67
\bibitem[Zheng et al. (2012b)]{zheng12b}
Zheng, R., Jiang, Y., Yang, J., et al.\ 2012b, \aap, 541, 49
\bibitem[Zheng et al. (2012c)]{zheng12c}
Zheng, R., Jiang, Y., Yang, J., et al.\ 2012c, \apj, 753, 112
\bibitem[Zheng et al. (2012d)]{zheng12d}
Zheng, R., Jiang, Y., Yang, J., et al.\ 2012d, \apjl, 753, 29
\bibitem[Zheng et al. (2013a)]{zheng13a}
Zheng, R., Jiang, Y., Yang, J., et al.\ 2013a, \apj, 764, 70
\bibitem[Zheng et al. (2013b)]{zheng13b}
Zheng, R., Jiang, Y., Yang, J., et al.\ 2013b, \mnras, 431, 1359
\bibitem[Zheng et al. (2014)]{zheng14}
Zheng, R., Jiang, Y., Yang, J., \& Erd\'{e}lyi, R.\ 2014, \mnras, 444, 1119
\bibitem[Zhou \& Liang (2017)]{zhou17}
Zhou, X. P., \& Liang, H. 2017, ChA\&A, 41, 224
\bibitem[Zong \& Dai (2015)]{zong15}
Zong, W., \& Dai, Y. 2015, \apj, 809, 151
\bibitem[Zong \& Dai (2017)]{zong17}
Zong, W., \& Dai, Y. 2017, \apjl, 834, L15
\bibitem[Zhang et al. (2012)]{zhang12}
Zhang, Q. M., Chen, P. F., Xia, C., \& Keppens, R. 2012, \aap, 542, A52
\bibitem[Zhang et al. (2017)]{zhang17}
Zhang, Q. M., Li, T., Zheng, R. S., Su, Y. N., \& Ji, H. S. 2017, \apj, 842, 27
\bibitem[Zhukov \& Auch\`{e}re(2004)]{zhuk04}
Zhukov, A. N., \& Auch\`{e}re, F. 2004, \aap, 427, 705
\bibitem[Zhukov et al.(2009)]{zhuk09}
Zhukov, A. N., Rodriguez, L., \& de Patoul, J. 2009, \solphys, 259, 73
\end{thebibliography}
\end{document}